\documentclass{article}
\usepackage{graphicx} % Required for inserting images
\usepackage{amsmath}
\usepackage{amsthm}

\usepackage{graphicx}
\usepackage{multicol}
\usepackage[version=4]{mhchem}
\usepackage[table,xcdraw]{xcolor}
\usepackage{float}

\usepackage{hyperref}
\usepackage{color}
\hypersetup{colorlinks=false, linktoc=all,linkcolor=blue}
\usepackage{caption}
\usepackage{subcaption}
\usepackage{verbatim}
\usepackage{mwe}
\usepackage[ruled,vlined]{algorithm2e}
\usepackage[english]{babel}
\usepackage{amsmath}
\usepackage{verbatim}

\usepackage[autostyle]{csquotes}
\usepackage{amsfonts}
\usepackage{amssymb}
\usepackage{float}
\usepackage{amsthm} 
\usepackage{amsmath}
\usepackage{enumitem}

\usepackage{lscape}

\usepackage{authblk}

\newtheorem{theorem}{Theorem}[section]
\newtheorem{definition}[theorem]{Definition}

\title{{\bf The Connection between Non-normality and Trophic Coherence in Directed Graphs}}
\author[1]{Catherine Drysdale}
\author[2]{Samuel Johnson}

\date{}

\affil[1]{Centre for Systems Modelling and Quantitative Biomedicine, University of Birmingham, UK}
\affil[2]{School of Mathematics, University of Birmingham, UK}

\begin{document}

\maketitle

\begin{abstract}
Trophic coherence and non-normality are both ways of describing the overall directionality of directed graphs, or networks. Trophic coherence can be regarded as a measure of how neatly a graph can be divided into distinct layers, whereas non-normality is a measure of how unlike a matrix is with its transpose. We explore the relationship between trophic coherence and non-normality by first considering the connections that exist in the literature and calculating the trophic coherence and non-normality for some toy networks. We then explore how persistence of an epidemic in an SIS model depends on coherence, and how this relates to the non-normality. A similar effect on dynamics governed by a linear operator suggests that it may be useful to extend the concept of trophic coherence to matrices which do not necessarily represent graphs.
\end{abstract}


\section*{Introduction}

%\color{{black}

In this perspective article, we aim to explore the relationship between trophic coherence and non-normality, which are both qualities used to describe directed graphs. Non-normality refers to the overall asymmetry of an adjacency matrix. Trophic coherence is defined as how neatly the network can be divided into distinct layers, but it can also be interpreted as a tendency of edges to align with a global direction. These two notions come together in directed graphs. A directed graph can be represented with an $N\times N$ adjacency matrix $A$. If the graph is unweighted then $A$ is binary: $A_{ij}=1$ if there is an edge from node $v_i$ to node $v_j$, else $A_{ij}=0$.
%\color{{blue}
When there is an edge from $v_i$ to $v_j$, we will say that $v_i$ ``sees'' $v_j$. A weighted directed graph can be represented with a matrix whose entries are real numbers. For our purposes here, we will always assume that $A$ is non-negative. 
%\color{{black}
Each node has an in-degree and an out-degree: $k_i^{in}=\sum_j A_{ji}$ and $k_i^{out}=\sum_j A_{ij}$. These are sometimes referred to as ``strengths'' if the directed graph is weighted. 
%We give the definitions of non-normality and trophic coherence below before presenting important notions and results in the literature before giving the structure of the rest of the article; 
\begin{definition}(Non-normality.)
    Given a real matrix $A$ and its transpose $A^{T}$, we say that the matrix $A$ is normal if $A A^T - A^{T}A = 0$. It is non-normal otherwise.
\end{definition} 
\begin{definition}{(Trophic Coherence)}
A directed graph is said to be maximally coherent if it is possible to assign to each node a natural number such that nodes assigned to $n$ only see
%connect to 
others assigned to $n+1$. The greater the deviation from such a configuration, the more incoherent the graph.
\end{definition}

The need to understand non-normal matrices and operators arises in the fields of Fluid Dynamics \cite{cossu1997global,chomaz2005global,sujith2016non,schmid2002stability}, PT-symmetry \cite{BenderComplexExtension,BenderNonHermitianHamiltoniansSense,BenderRealSpectrainNon-HermitianHamiltonians,BenderHamiltonianHermitian,CarlMBenderHiddenSymmetry,Mostafazadeh_2006,PhysRevD.86.121702,DaviesAnharmonic} and Mathematical Biology \cite{drysdale2024novel} amongst other disciplines.
In particular, non-normal systems are often characterised with eigenvalues that are sensitive to perturbation. It is for this non-normality is considered an asset in information transfer and communication as it can amplify small environmental changes \cite{baggio1, baggio2}. Additionally, non-normal linear operators are difficult to capture numerically as small discrepancies arising from machine precision can manifest as large perturbations in the eigenvalues. This has given rise to the study of pseudospectra; the $\epsilon$-pseudospectrum of a matrix $A$ is the defined as the set $\{ z \in \mathbb{C} : || (A - zI)^{-1} || < \epsilon^{-1} \}$). The definition of pseudospectra corresponds to a measure of sensitivity to perturbations of size $\epsilon$; an equivalent definition of pseudospectra is the set $\{ z \in Sp(A + B) : || B || \leq \epsilon \}$ \cite{trefethen2005spectra}. Despite this being one of the most useful tools for understanding how eigenvalues respond to perturbation, it is not
%so much of a useful tool 
necessarily the best way of
%for 
understanding the sensitivity of directed graphs to operators such as changing the weights of edges, or edge deletion. This is equivalent to putting a structure on the matrix $B$, which thus breaks the correspondence between the ``perturbation view of pseudospectra" and the ``transient phenomena" view of pseudospectra. The latter is particularly relevant when the adjacency matrix represents a discrete dynamical system.

%%\color{{blue}
%Trophic coherence has its background in ecology, and ecologists usually
%%\color{{black}
Ecologists define the ``trophic level'' of a species as the average level of its prey, plus one \cite{Levine_levels}. Trophic coherence was first proposed as a solution to May's paradox: the fact that large ecosystems are stable \cite{johnson2014trophic}. If the trophic difference of each edge in a graph is the difference between the trophic levels of the in- and out-neighbours, then the broader the distribution of differences (i.e. the larger the standard deviation of this distribution), the more incoherent the network. %\color{{black}
It was found that ecosystem models based on sufficiently coherent graphs became more stable, rather than less so, with increasing size. It was subsequently shown, by means of graph ensembles and numerical simulations, that trophic coherence could be related to several aspects of directed networks more generally, including the spectral radius and distribution of cycles \cite{Johnson_looplessness}; motif profiles \cite{Janis_motifs}; non-normality and strong connectivity \cite{johnson2020digraphs,rodgers2023strong}; pseudospectra \cite{rodgers2023influence}; and various dynamical processes \cite{klaise2016neurons,pilgrim2020organisational,rodgers2022network}. However, relying on the ecological definition of trophic levels restricted the application of trophic coherence to networks with at least one node with in-degree zero. So a new measure of trophic
%\color{{blue}
levels
%\color{{black}
was proposed that can be applied to any directed graph \cite{HowDirected}.
%\color{{blue}
This is the method we use here.

In this article, we wish to emphasize the connection between trophic coherence and non-normality; and to suggest that this may be relevant not only for directed graphs, but for other systems described by matrices. In the first section, ``Measuring Trophic Coherence and Non-normality", we take a deeper look into the literature and present results that connect the two ideas. We also calculate the non-normality and trophic coherence of some toy networks in order to help the reader build an intuition. In the second section, we study an SIS model and a simple linear dynamics, both of which are affected by the coherence of an underlying matrix. We see how non-normality, strong connectivity and the spectral radius are also  determined by the trophic coherence. We then conclude by discussing potential further avenues of research to establish stronger bonds between these topics. 

\section*{Measuring trophic coherence and non-normality}

\begin{definition}
The vector of trophic levels ${\bf h}$ of a directed graph with adjacency matrix $A$ is the solution to the equation
\begin{equation}
\Lambda {\bf h} = {\bf v},
\label{eq_h}
\end{equation}
where
\begin{equation}
\Lambda=\mbox{diag}({\bf u})-A-A^T,
\label{eq_Lambda}
\end{equation}
and the vectors of total degree and degree imbalance are, respectively, ${\bf u}={\bf k^{in}}+{\bf k^{out}}$ and ${\bf v}={\bf k^{in}}-{\bf k^{out}}$ \cite{HowDirected}. Since the solution to Eq. (\ref{eq_h}) is defined only up to an additive constant, the convention that $\mbox{min}(h_i)=0$ is used (i.e. all the elements of ${\bf h}$ are positive except for the smallest value which is set to zero.)
\label{def_h}
\end{definition}

While Eq. (\ref{eq_h}) always has a solution, which is unique given the convention stated at end of the definition, we should note that one cannot obtain this solution by inverting $\Lambda$, since this matrix is always singular. One must therefore use some other method to find the solution, such as LU decomposition, the Moore-Penrose pseudo-inverse, or an iterative method.

We can measure the trophic coherence of a directed graph with the {\it incoherence parameter} $F$, given by:
\begin{definition}
The trophic incoherence of a directed graph with adjacency matrix $A$ and trophic levels ${\bf h}$ given by Eq. (\ref{eq_h}) is 
\begin{equation}
F = \frac{\sum_{ij} A_{ij}(h_j - h_i - 1)^2}{\sum_{ij}A_{ij}}.
\label{eq_F}
\end{equation} 
\label{def_F}
\end{definition}

The incoherence $F$ would coincide with the square of the parameter $q$ proposed by \cite{johnson2014trophic} if trophic levels were calculated as in ecology
%\color{{blue}
($F=q^2$)
\cite{johnson2014trophic}.
Using the trophic levels given by Eq. (\ref{eq_h}),
%\color{{black}
$F$ is bounded between zero and one: $F=0$ implies a perfectly coherent directed graph in which vertices fit into integer trophic levels; and $F=1$ corresponds to maximum incoherence, which occurs if and only if the directed graph is balanced (${\bf v}={\bf 0}$) \cite{HowDirected}.

%\color{{blue}
Definitions \ref{def_h} and \ref{def_F} were originally derived by first writing down Eq. (\ref{eq_F}) for generic levels, and then finding the solution ${\bf h}$ that minimised $F$
%by taking partial derivatives with respect to $h_i$ and equating to zero 
-- which leads to Eq. (\ref{eq_h}). In other words, the trophic levels, under this definition, are those which minimise the trophic incoherence.

%\color{{black}

The average trophic difference is $z=1-F$, so one interpretation of trophic coherence is as the ``directedness'' of the graph,
%\color{{blue}
and $z$ can be referred to as the trophic coherence
\cite{HowDirected}.
%\color{{black}
Another interpretation is given by SpringRank \cite{de2018physical}, which likens each edge to a spring with natural length $l=1$. $F$ is then the energy, which is minimised at the solution ${\bf h}$. Yet another interpretation of the same equation is Helmholtz-Hodge decomposition, whereby a vector field can be decomposed into a gradient part and a zero divergence part \cite{kichikawa2019community}. When applied to graphs, $F$ is then its ``circularity''. The fact the same equation has appeared independently at least three times testifies to its wide applicability \cite{HowDirected}.
%\color{{blue}
We note also that if the ``$-1$'' in Eq. (\ref{eq_F}) were replaced with a constant ``$-a$'', this would simply multiply the trophic levels by $a$: ${\bf h'}=a{\bf h}$. So $a=1$ is a natural choice that does not reduce generality.
%\color{{black}

%\color{{blue}
$\Lambda$ is twice the Laplacian of the undirected version of the graph $(A + A^{T})/2$, which can be considered an undirected graph as all edges now have an opposite edge of the same weight. In particular, an interpretation is that $\Lambda {\bf h} = {\bf v}$ is the corresponding inhomogeneous equation to the homogeneous equation $(L(A) + L(A^{T})) {\bf x} = 0 $. The multiplicity of the eigenvalue $0$ corresponds to the number of connected components in the undirected case, hence \ref{eq_Lambda} can be seen as when we ``force" $(L(A) + L(A^{T})) {\bf x} = 0$, by unbalancing the in and out degrees on each node.

%%\color{{red}
%To be edited:
%%\color{{black}

%\color{{blue}
Whereas the trophic coherence can be captured by a single number, it is not so easy to have a single number which captures non-normality.
%\color{{black}
Various measures have been proposed to quantify the non-normality of matrices, the most obvious being $|| AA^{T} - A^{T}A||$ for some suitable norm (here and in the paper we use the Frobenius norm). Another method is Henrici's deviation from normality: $Hen(A) = \sqrt{||A||_{F}^{2} - \sum_{i=1}^{n} |\lambda_i|^2}$, where $||\cdot||_{F}$ is the Frobenius norm and $\lbrace \lambda_i\rbrace$ are the eigenvalues of $A$. This can be normalised by dividing through by the $||A||_{F}^{2}$
%\color{{blue}
giving the parameter
\begin{equation}
    d_F=\sqrt{1-\nu}, \hspace{1cm} \text{where}\hspace{1cm} \nu=    \frac{\sum_{i=1}^{n} |\lambda_i|^2}{||A||_{F}^{2}}
    \label{eq_df}
\end{equation} is known as the ``normality'' \cite{HowDirected}.
Note that a directed graph with normal adjacency matrix $A$ would have $\sum_{j=1}^N |\lambda_j|^2 = ||A||^2_F $ and hence $\nu = 1$ ($d_F=0$); but a ``very non-normal" network would have $|\lambda_j| =0$ for all $j$ and in this case $\nu = 0$ ($d_F=1$). However, we stress that the idea of ``very non-normal" is also subjective with respect to which measure. The norm $|| AA^{T} - A^{T}A||$ can be considered more sensitive to structural features such as skewness or asymmetry, whereas Henrici's measure aggregates deviations related to eigenvalue magnitudes, potentially smoothing out localised anomalies. In particular, for matrices close to being symmetric, but not normal, the two measures might diverge significantly. The measure $|| AA^{T} - A^{T}A||$ could show a large deviation, while Henrici's measure may remain small if the eigenvalues are unaffected.

In Table 1, we compute the trophic coherence and non-normality measures for two graphs (the loop on five vertices and the so-called vortex graph on five vertices). We consider these measures for both the adjacency matrix and the non-symmetric Laplacian $L(A)$. We also add an edge to the cycle and delete an edge. We see that trophic coherence increases (trophic incoherence $F$ decreases) in both settings. Also, we can relate back to our previous discussion regarding pseudospectra. where deletion and addition of an edge can be considered a norm of the same size hence a more nuanced approach is needed. Also, we have that the two measures of non-normality behave differently as the Henrici norm perceives the graph with edge deletion as more non-normal regarding the adjacency matrix and the Laplacian, whereas the converse is true for the other measure. Further analysis is needed to establish if there is a physical meaning to this in certain scenarios, i.e. the graph represents a dynamical system. Although, in both cases the orientation of each edge can be reversed resulting in the same graph, the fact that flow is not conserved on the ``loop under edge addition" may make this graph more asymmetric. Further work must be done in this direction to establish this. 

Asllani {et al.}
\cite{asllani2018structure} have recently studied %\color{{black}
the effects of non-normality in directed graphs, where it was shown that
%\color{{blue}
$d_F$
%\color{{black}
correlates strongly with a measure of structural asymmetry. Furthermore, the structures of graphs with different degree distributions were calculated. The approach of the Henrici norm has also been used recently in work looking at non-normality in the context of trophic coherence \cite{johnson2020digraphs,HowDirected}. 

It is possible to estimate the expected value of various magnitudes given that a network has a specified trophic coherence, by means of graph ensembles \cite{Johnson_looplessness,johnson2020digraphs,HowDirected}. The ``coherence ensemble'' is the set of all possible directed graphs with given degree sequence and trophic coherence. Thus, the expected value of the spectral radius is
\begin{equation}
\overline{\rho}=e^\tau, \hspace{1cm} \text{where} \hspace{1cm} \tau=\ln \alpha + \frac{L_B}{2(L-L_B)} - \frac{1-F}{2F},
    \label{eq_rho}
\end{equation}
and the bar denotes expectation \cite{Johnson_looplessness}. $L$ is the number of edges, $L_B$ the number of edges connected to source or sink nodes (those with no incoming or outgoing edges), and the ``branching factor'' is $\alpha= \langle k^{in} k^{out}\rangle / \langle k\rangle$ where the brackets are averages over vertices. $\tau$ is referred to as the ``loop exponent''. Because $\tau$ is positive for $F\simeq 1$ but becomes negative when $F\simeq 0$, directed graphs fall into one of two regimes, referred to as ``loopful'' ($\tau>0$) and ``loopless'' ($\tau<0$) \cite{Johnson_looplessness}. In the former the number of circuits of length $l$ increases exponentially with $l$, whereas in the latter they decay exponentially. This can have a crucial bearing on many other topological and dynamical features of complex systems, as we illustrate below with the example of an SIS model: an epidemic perdures indefinitely when $\tau>0$ but quickly goes extinct when $\tau<0$.

This approach has been used to show that the expected non-normality is bounded, $\overline{d_F}\geq\sqrt{1-e^{2\tau}/\langle k\rangle}$ \cite{johnson2020digraphs}, and approximated by $\overline{d_F}\simeq \sqrt{1-\exp(1-1/F)}$ \cite{HowDirected}. Hence, trophically coherent graphs ($F\rightarrow 0$ or $\tau\rightarrow-\infty$) are non-normal
%\color{{blue}
($d_F \rightarrow 1)$.
The fact that the expected value of the non-normality is bounded below by the expected spectral radius \cite{johnson2020digraphs} is natural when considering the transient wave-packet phenomena that can happen with non-normal matrices \cite{trefethen2005spectra}.

%%%%%%%%%%%%%%%%%%%%%%%%%%%%%%%%%%%%%%%%%%%%%%%%%%%%%%%%%%%%%%%%%%%%%%%%%%%%%%%%%%%%%%%%%%%%%%%%%%%%%%%%%%%%%%%%%%%%%%%%%%%%%%%%%%%%%%%%%%%%%%%%%%%%%%%%%%
%\section*{Example: SIS model on coherent networks}

\begin{figure}[ht]
%\begin{center}
\centering
\includegraphics[width=11cm]{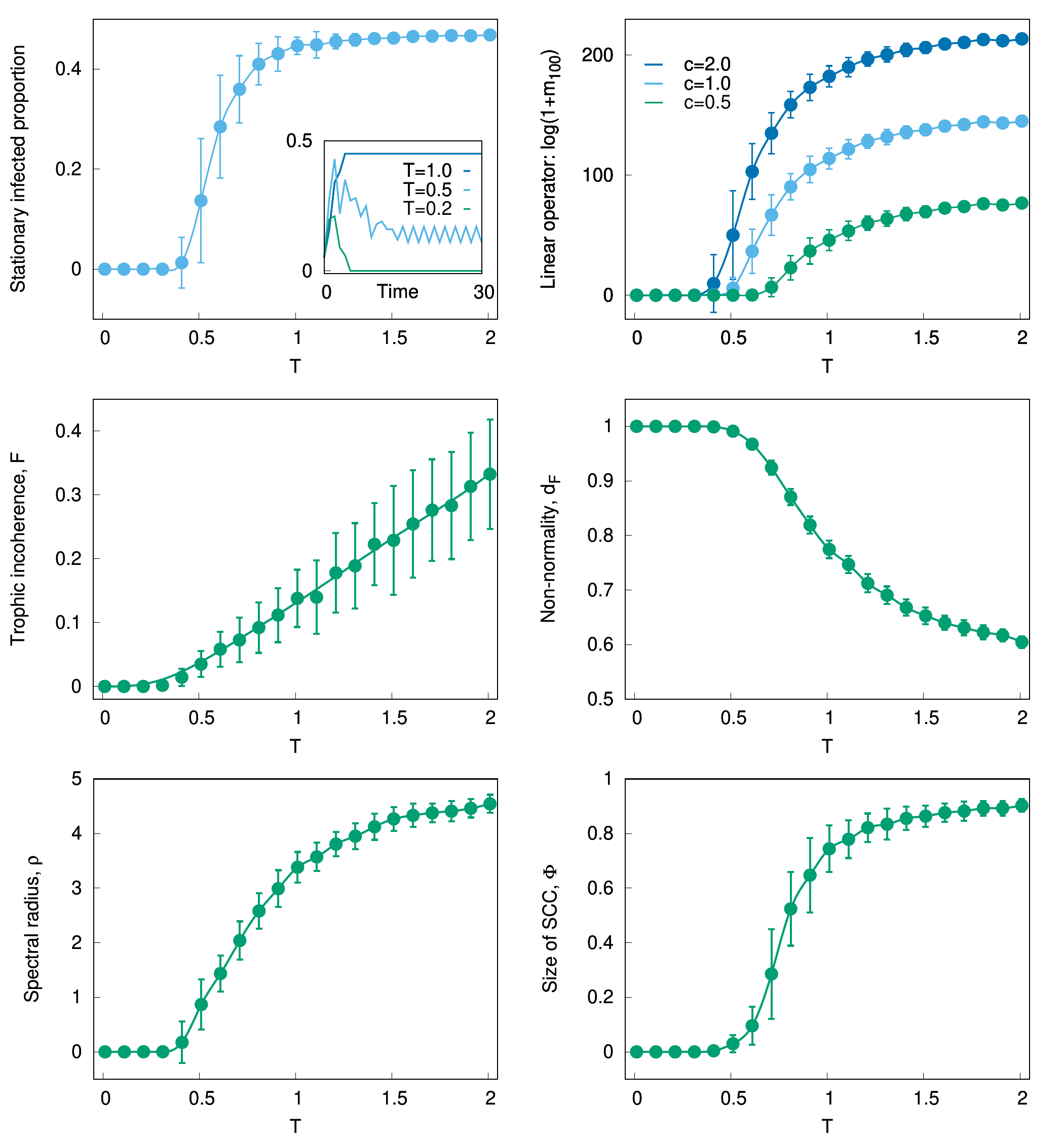}%{fig_op_all.eps} % Ensure the .eps file is in the same directory or adjust the path
%\end{center}
\caption{
    {\bf Top left:} Stationary proportion of infected nodes against parameter $T$ for networks generated with the Generalised Preferential Preying Model \cite{klaise2016neurons}. Number of nodes and edges: $N=100$ and $L=500$. Averages over 100 networks; error bars are standard deviations. Stationary values computed as the average from $t=90$ to $t=100$. Inset: Time series of the proportion of infected nodes for three networks, generated with $T=1$, $0.5$, and $0.2$.
    {\bf Top right:} Mean activity $m(t)$ at $t=100$ according to Eqs. (\ref{eq_xt}) and (\ref{eq_mt}), where $A$ is the adjacency matrix of the networks used in the top left panel. Initially all elements have $x(0)=0$, except for a randomly chosen $5\%$ which are set to $x(0)=1$. The extent of activity is measured as $\ln[1+m(100)]$, for $c=0.5$, $1$, and $2$.
    {\bf Middle left:} Trophic incoherence $F$ against $T$ for the same directed graphs as in the panels above.
    {\bf Middle right:} Non-normality $d_F$ against $T$ for the same directed graphs.
    {\bf Bottom left:} Spectral radius $\rho$ against $T$ for the same directed graphs.
    {\bf Bottom right:} Size of the strongly connected component ${\Phi}$ against $T$ for the same directed graphs.
}
\label{fig_epi}
\end{figure}

\section*{Spreading processes with graphs and operators}

%\color{{blue}
It is known that the trophic coherence of directed graphs can exert an important influence on the dynamics of various complex systems \cite{johnson2020digraphs}. Likewise, in dynamical systems governed by linear operators, the non-normality thereof will fundamentally affect system behaviour \cite{trefethen2005spectra}. We go on to show, using two simple examples of spreading processes, that there are close similarities between these two kinds of phenomena. First we look at the SIS model on coherent directed graphs, and then we compare this to the action of a non-normal linear operator.
These examples also show that the methods which have been proposed to generate directed graphs with tunable trophic coherence might also be useful for studying other matrices numerically \cite{klaise2016neurons,rodgers2024fitness}.

%\color{{black}
The SIS (Susceptible-Infectious-Susceptible) model is perhaps the simplest which can be used to study the spread through a population of an infectious disease -- typically one which confers little or no immunity \cite{hethcote1989three}. We will use this paradigm to demonstrate the influence of trophic coherence or non-normality on even the simplest of dynamical processes.

Consider an unweighted, directed network given by the $N\times N$ adjacency matrix $A$. To each node $i$ is associated a dynamical variable $s_i(t)$ which can take, at each discrete time $t$, either the value $0$ or $1$, representing susceptible or infectious states, respectively. Let $g_i(t)=\sum_j A_{ji}s_j(t)$. The system then evolves according to:
\begin{align}
s_i(t+1)&=1\qquad\mbox{ if } s_i(t)=0\mbox{ and }g_i(t) > 0, \mbox{or}\\
    s_i(t+1)&=0 \qquad \mbox{otherwise,}
\end{align}
with all nodes updated in parallel. In other words, a susceptible node becomes infectious for one time step if at least one of its in-neighbours is infectious.

Will an epidemic die out naturally or go on indefinitely? We study this by beginning with all nodes being susceptible except for $\%5$ which are chosen at random to be made infectious. We generate networks using the Generalised Preferential Preying Model \cite{klaise2016neurons}. This model has a parameter, $T$, which allows one to set the trophic coherence of the network: $T=0$ produces maximally coherent structures, and incoherence increases with positive $T$.

Not that in this scheme the only randomness is in the generation of the network and the choice of initial conditions; the dynamics thereafter is deterministic. Fig. \ref{fig_epi} (top left) shows the stationary proportion of nodes which remain infectious indefinitely against $T$. At low values of $T$ the epidemic dies out, but for higher values there is a continuous transition to a regime in which a significant proportion of the nodes remain infectious. This can be understood by considering that the epidemic requires a strongly connected component of nodes to sustain itself, which only exists in sufficiently incoherent networks \cite{rodgers2023strong}. Or, more formally, it is known that the critical rate of infection required for an epidemic to survive is lower bounded by the inverse of the spectral radius of the adjacency matrix \cite{van2013non}, which depends on trophic coherence \cite{Johnson_looplessness}.

%We see that growth in spectral radius coincides with the decrease in non-normality, but this could also be because the lead eigenvalue is increasing with the increased $T$  Fig. \ref{fig_epi} (top left).
%\color{{blue}
As we have seen, non-normality also varies with trophic coherence.
Hence, in the inset we see a ``bump'' in the time series for lower values of $T$ (which produces small $F$ and high $d_F$), which corresponds to transient phenomena. In a future work, we will consider the average difference between the pseudospectral abscissa and the spectral radius at early times to establish transient phenomena \cite{trefethen2005spectra}.
In our SIS model computations, we see that as trophic coherence increases, the size of the strongly-connected component increases. Unlike in undirected graphs, the multiplicity of the the $\lambda=0$ eigenvalue of the Laplacian $L(A)$ does not equal the number of components, but rather the number of reaches \cite{veerman2020primer} (a reach is the maximal unilaterally connected set). 
%If a graph has more than one maximal reach, then it must have the same number of disconnected components (an edge between disconnected components would increase the size of the reach but nevertheless there would now only be one reach instead of two). The number of disconnected components is an upper bound on the size of size of strongly connected components.
Whereas dynamical processes such as these have been related to the existence and size of strongly connected components, we have yet to investigate the effect of different reach structures.
%We have not yet explored the significance of the multiplicity of the $\lambda=0$ eigenvalue of $L(A) + L(A^{T})$, which is the operator that determines trophic coherence as well as the Laplacian of the corresponding undirected graph. However, as non-normality can manifest in degeneracy of eigenvalues, in a future work, we will explore the multiplicity of the eigenvalue $0$ of $L(A)$ and $L(A) + L(A^{T})$ as the temperature increased as this would be a complimentary measure to the size of the strongly-connected component.

% New bit on linear operator:
%\color{{blue}
Consider now the following dynamical system with $N$ elements. Every element $i$ is characterised by a continuous dynamical variable $x_i(t)$ at discrete time $t$. The system evolves according to
\begin{equation}
    {\bf x}(t+1)= c A {\bf x}(t),
    \label{eq_xt}
\end{equation}
where $c$ is a constant parameter and $A$ is a non-negative, $N\times N$ real matrix. We might consider $A$ as a linear operator or as a directed graph on which the process is taking place. In particular, we can take $A$ to be infinite dimensional in which case we could be giving a graphical interpretation to such linear operators. We begin with a small number of randomly chosen agents in state $x(0)=1$ and all others $x(0)=0$, and track the average value of the activity, 
\begin{equation}
m(t)=\frac{1}{N}\sum_{i=1}^N x_i(t).
\label{eq_mt}
\end{equation}
This will either decay to zero or diverge according to the spectral radius of $A$ and the parameter $c$, as shown in Fig. \ref{fig_epi} (top right). However, for significantly non-normal networks (low $T$)
there may be transient behaviour that eventually dies out, as in the SIS case on a trophically coherent graph. This might correspond to an epidemic, rumour or other spreading process which reaches most of the system but goes on to disappear; whereas in the more normal or incoherent case, even a process that only reaches some of the system might continue to fuel itself indefinitely thanks to feedback.  
%the average activity is small. 
%Therefore, it could be argued that the non-normality is compensated by a larger spectral radius. A metaphor could be that a disease which is infectious over longer distances (in this case large spectral radius) can fair better even when individuals are separated. Additionally, we do not know how the non-normality of the underlying graph manifests as if it is a higher multiplicity of $0$ (completely disconnected components), then it could be that the epidemic would be incapable of spreading. 

%We can regard $A$ as the operator defining the dynamics; or as the adjacency matrix of a network on which the dynamics is taking place, with the variables at the nodes. Thus, this model provides a simple test case allowing for the comparison of concepts such as trophic coherence and non-normality as they pertain both to operators and to graphs.

%\color{{black}

Fig. \ref{fig_epi} (middle panels) shows how the trophic incoherence $F$ and the non-normality $d_F$ also vary in this network model with the parameter $T$. A comparison with Fig. \ref{fig_epi} (top left) reveals that some nodes begin to sustain the epidemic once $F>0$ or $d_F<1$;
%\color{{blue}
and a similar effect is evident in the linear operator case (top right).
%\color{{black}
Fig. \ref{fig_epi} (bottom panels) shows how the two topological features we can relate to
%\color{{blue}
both the SIS dynamics
and the linear operator
%\color{{black}
-- namely, the spectral radius $\rho$ and the size of the strongly connected component $\Phi$ - undergo a similar transition with increasing $T$ as the stationary proportion of infected nodes
%\color{{blue}
or the logarithm of $m(t)$.
%\color{{black}
Just as degree heterogeneity can drastically reduce the size of epidemic waves \cite{johnson2024epidemic}, trophic coherence can affect their extinction.
%\color{{blue}
And whereas trophic coherence has to date been thought of only as a property of directed graphs, this example suggests that it can be studied in the case of operators and square matrices more broadly.
%\color{blue}

%%%%%%%%%%%%%%%%%%%%%%%%%%%%%%%%%%%%%%%%%%%%%%%%%%%%%%%%%%%%%%%%%%%%%%%%%%%%%%%%%%%%%%%%%%%%%%%%%%%%%%%%%%%%%%%%%%%%%%%%%%%%%%%%%%%%%%%%%%%%%%%%%%%%%%%%%%%%%%%%%%%%%%%%%%%%%%%%%%%%%%%%%%%%%%%%%%%%%%%%%%%%%%%%%%%%%%%%%%%%%%%%%%%%%%%%%%%%%%%%%%%%%%%%%%%%%%%%%%%%%%%%%%%%%%%%%%%%%%%%%%%%%%%%%%%%%%%%%%%%%%%%%%%%%%%

%%%%%%%%%%%%%%%%%%%%%%%%%%%%%%%%%%%%%%%%%%%%%%%%%%%%%%%%%%%%%%%%%%%%%%%%%%%%%%%%%%%%%%%%%%%%%%%%%%%%%%%%%%%%%%%%%%%%%%%%%%%%%%%%%%%%%%%%%%%%%%%%%%%%%%%%%%%%%%%%%%%%%%%%%%%%%%%%%%%%%%%%%%%%%%%%%%%%%%%%%%%%%%%%%%%%%%%%%%%%%%%%%%%%%%%%%%%%%%%%%%%%%%%%%%%%%%%%%%%%%%%%%%%%%%%%%%%%

%\color{black}

\section*{Discussion}

%\color{black}
In this article, we have discussed the relationship between trophic coherence and non-normality. We have mentioned some existing connections in the literature and studied some small graphs for illustration.
We have also presented numerical experiments in the form of an SIS model and a linear dynamics, which show how trophic coherence and non-normality are related and have significant effects on dynamical systems governed by matrices. 

There are many relationships still to be discovered, particularly regarding the connection between non-normality and trophic coherence in matrices in general. Also of interest is the relationship with strongly-connected components, and how edge deletion and edge addition may be considered in a way that is amenable to the calculation of pseudospectra. In particular, we aim to explore which kind of edge perturbations can create the largest change in non-normality, trophic coherence, or even other measures such as algebraic connectivity (which is non-trivial to compute in a directed graph and is so far yet to profit from the advances in the computations of non-self-adjoint problems). By studying how such edge perturbations change the non-normality, trophic coherence and algebraic connectivity, we may improve our understanding of how such magnitudes are related, and their effects on dynamical systems. This is the subject of upcoming work.

%\subsubsection*{Conflict of Interest Statement}
\paragraph{Conflict of Interest Statement:}
The authors declare that the research was conducted in the absence of any commercial or financial relationships that could be construed as a potential conflict of interest.
%\color{{black}

%\subsubsection*{Author Contributions}
\paragraph{Author Contributions:}
The two authors contributed to this manuscript equally. 

%\subsubsection*{Acknowledgments}
\paragraph{Acknowledgments: }
The authors would like to thank the Centre for Systems Modelling and Quantitative Biomedicine, University of Birmingham, for ``seedcorn'' funding; as well as Vagabond in Birmingham and Black Sheep Coffee for providing wine and Lion's mane lattes respectively.

\begin{landscape}
\begin{table}
\centering
\begin{tabular}%{|p{2.75cm}|p{4.75cm}|p{4.75cm}|p{4.75cm}|p{4.75cm}|} 
{|p{2.5cm}|p{4.5cm}|p{4.5cm}|p{4.5cm}|p{4.5cm}|} 
\hline
Property 
&   \parbox{4.5cm}{{Vortex Graph on five vertices $V_5$ } \\ \includegraphics[scale=0.5]{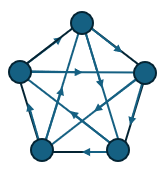}}
&   \parbox{4.5cm}{{Loop on five vertices $P_5$} \\ \includegraphics[scale=0.5]{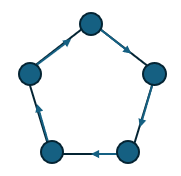}}
&   \parbox{4.5cm}{{Loop under edge deletion (chain)} \\ \includegraphics[scale=0.5]{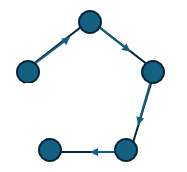}}
&   \parbox{4.5cm}{{Loop under edge addition } \\ \includegraphics[scale=0.5]{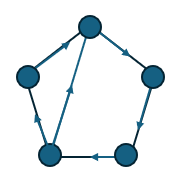}}
\\
\hline
Adjacency Matrix &
\small
\vspace{0.3cm}
$\begin{pmatrix} 
0 & 1 & 1 & 0 & 0 \\ 
0 & 0 & 1 & 1 & 0\\ 
0 & 0 & 0 & 1 & 1\\ 
1 & 0 & 0 & 0 & 1 \\
1 & 1 & 0 & 0 & 0 
\end{pmatrix}$
\vspace{0.3cm}
& 
\small
\vspace{0.3cm}
$\begin{pmatrix} 
0 & 1 & 0 & 0 & 0 \\ 
0 & 0 & 1 & 0 & 0\\ 
0 & 0 & 0 & 1 & 0\\ 
0 & 0 & 0 & 0 & 1 \\
1 & 0 & 0 & 0 & 0 
\end{pmatrix}$
\vspace{0.3cm}
&
\small
\vspace{0.3cm}
$\begin{pmatrix} 
0 & 1 & 0 & 0 & 0 \\ 
0 & 0 & 1 & 0 & 0\\ 
0 & 0 & 0 & 1 & 0\\ 
0 & 0 & 0 & 0 & 1 \\
0 & 0 & 0 & 0 & 0 
\end{pmatrix}$
\vspace{0.3cm}
&
\small
\vspace{0.3cm}
$\begin{pmatrix} 
0 & 1 & 0 & 0 & 0 \\ 
0 & 0 & 1 & 0 & 0\\ 
0 & 0 & 0 & 1 & 0\\ 
0 & 0 & 0 & 0 & 1 \\
1 & 1 & 0 & 0 & 0 
\end{pmatrix}$
\vspace{0.3cm}\\ 
\hline
\parbox{3cm}{Laplacian\\($L = D^{out} - A$) }
&
\small
\vspace{0.3cm}$\begin{pmatrix} 
2 & -1 & -1 & 0 & 0 \\ 
0 & 2 & -1 & -1 & 0\\ 
0 & 0 & 2 & -1 & -1\\ 
-1 & 0 & 0 & 2 & -1 \\
-1 & -1 & 0 & 0 & 2 
\end{pmatrix}$
\vspace{0.3cm}
& 
\small
\vspace{0.3cm}
$\begin{pmatrix} 
1 & -1 & 0 & 0 & 0 \\ 
0 & 1 & -1 & 0 & 0\\ 
0 & 0 & 1 & -1 & 0\\ 
0 & 0 & 0 & 1 & -1 \\
-1 & 0 & 0 & 0 & 1 
\end{pmatrix}$
\vspace{0.3cm}
&
\small
\vspace{0.3cm}
$\begin{pmatrix} 
1 & -1 & 0 & 0 & 0 \\ 
0 & 1 & -1 & 0 & 0\\ 
0 & 0 & 1 & -1 & 0\\ 
0 & 0 & 0 & 1 & -1 \\
0 & 0 & 0 & 0 & 0 
\end{pmatrix}$
\vspace{0.3cm}
&
\small
\vspace{0.3cm}
$\begin{pmatrix} 
1 & -1 & 0 & 0 & 0 \\ 
0 & 1 & -1 & 0 & 0\\ 
0 & 0 & 1 & -1 & 0\\ 
0 & 0 & 0 & 1 & -1 \\
-1 & -1 & 0 & 0 & 2 
\end{pmatrix}$
\vspace{0.3cm}\\ 
\hline
\raggedright
Trophic Incoherence ($F$) &  1 &  1 &  0 & 0.91 \\ 
\hline
Trophic Levels & (0,0,0,0,0) & (0,0,0,0,0) & (0,1,2,3,4)& (0,0.18, 0.27, 0.36, 0.54)  \\
\hline
Adjacency Non-normality $|| A^{T}A - AA^{T}||$ & 0 & 0 & 1.41 & 2.44\\ 
\hline
Laplacian Non-normality $|| L^{T}L - LL^{T}||$ & 0 & 0 & 2 & 4 \\
\hline
Adjacency Non-normality (Henrici) & 0 & 0 & 1 & 0.37 \\ 
\hline
Laplacian Non-normality (Henrici) & 0 & 0 & 0.71 & 0.38 \\ \hline
\end{tabular}
\caption{Comparison of various properties for different types of networks (Vortex Graph on 5 vertices, Loop on 5 Vertices, Loop under Edge Addition, Loop Under Edge Deletion).}
\end{table}
\end{landscape}
%\color{{black}

%\bibliographystyle{Frontiers-Harvard} 
%\bibliographystyle{ieeetr}
%\bibliographystyle{unsrtnat}
%\bibliographystyle{Frontiers-Vancouver}
%FrontiersinVancouver
%\bibliography{test}

\end{document}